 \documentclass[aps,prb,notitlepage,superscriptaddress,twocolumn ]{revtex4}
\usepackage{graphicx,subfigure,epsfig}
\usepackage{dcolumn}
\usepackage{amssymb}
\usepackage{times}
\usepackage{amsmath}
\usepackage{amsfonts}
\usepackage{mathrsfs}
\usepackage{setspace}
\usepackage{latexsym}
\usepackage{bbm}
\usepackage{float}
\usepackage{flafter}
\usepackage{bm}
\usepackage{epstopdf}
\usepackage{hyperref}
\usepackage{color}
\usepackage{multirow}

\def \be {\begin{eqnarray}}
\def \ee {\end{eqnarray}}

\begin{document}

\title{Valley Stoner Instability of the Composite Fermi Sea}

\author{Zheng Zhu}
\affiliation{Department of Physics, Massachusetts Institute of Technology, Cambridge, MA, 02139, USA}
\author{D. N. Sheng}
\affiliation{Department of Physics and Astronomy, California State University, Northridge, CA, 91330, USA}
\author{Liang Fu}
\affiliation{Department of Physics, Massachusetts Institute of Technology, Cambridge, MA, 02139, USA}
\author{Inti Sodemann}
 
\affiliation{Max-Planck Institute for the Physics of Complex Systems, D-01187 Dresden, Germany}

\begin{abstract}
We study two-component electrons in the lowest Landau level at total filling factor $\nu _T=1/2$ with anisotropic mass tensors and principal axes rotated by $\pi/2$ as realized in Aluminum Arsenide (AlAs) quantum wells. Combining exact diagonalization and the density matrix renormalization group we demonstrate that the system undergoes a quantum phase transition from a gapless state in which both flavors are equally populated to another gapless state in which all the electrons spontaneously polarize into a single flavor beyond a critical mass anisotropy of {\bf $m_x/m_y \sim 7$}. We propose that this phase transition is a form of itinerant Stoner transition between a two-component and a single-component composite fermi sea states and describe a set of trial wavefunctions which successfully capture the quantum numbers and shell filling effects in finite size systems as well as providing a physical picture for the energetics of these states. Our estimates indicate that the composite Fermi sea of AlAs is the analog of an itinerant Stoner magnet with a finite spontaneous valley polarization. We pinpoint experimental evidence indicating the presence of Stoner magnetism in the Jain states surrounding $\nu=1/2$.
\end{abstract}

\maketitle

\section{Introduction.} Since its discovery more than three decades ago~\cite{Tsui1982,Laughlin83} the fractional quantum Hall (FQH) regime has gifted us with a remarkably rich arena for correlated phases of two-dimensional (2D) electrons. Among these phases, a prominent example is the fractionalized gapless state proposed by Halperin, Lee, and Read~\cite{HLR} as a Fermi liquid-like state of the composite fermions (CFs) introduced by Jain~\cite{JainBook,Jain}. When the lowest Landau level has two nearly degenerate components, as it is the case of Gallium Arsenide (GaAs) in the limit of small $g$-factor, experimental~\cite{Du95,Kukushkin99,Tiemann2012} and numerical studies~\cite{Park2001} indicate that the composite fermi sea is spin un-polarized, namely, that the interactions are unable to induce a Stoner-like instability into a polarized state. In fact, systems of itinerant fermions for which the Stoner state can be theoretically established to be the ground state in an unbiased fashion are rare, typically because the Stoner instability appears at strong coupling where it is hard to rule out other competing correlated states.

\begin{figure}[btp]
\begin{center}
\includegraphics[width=0.4\textwidth]{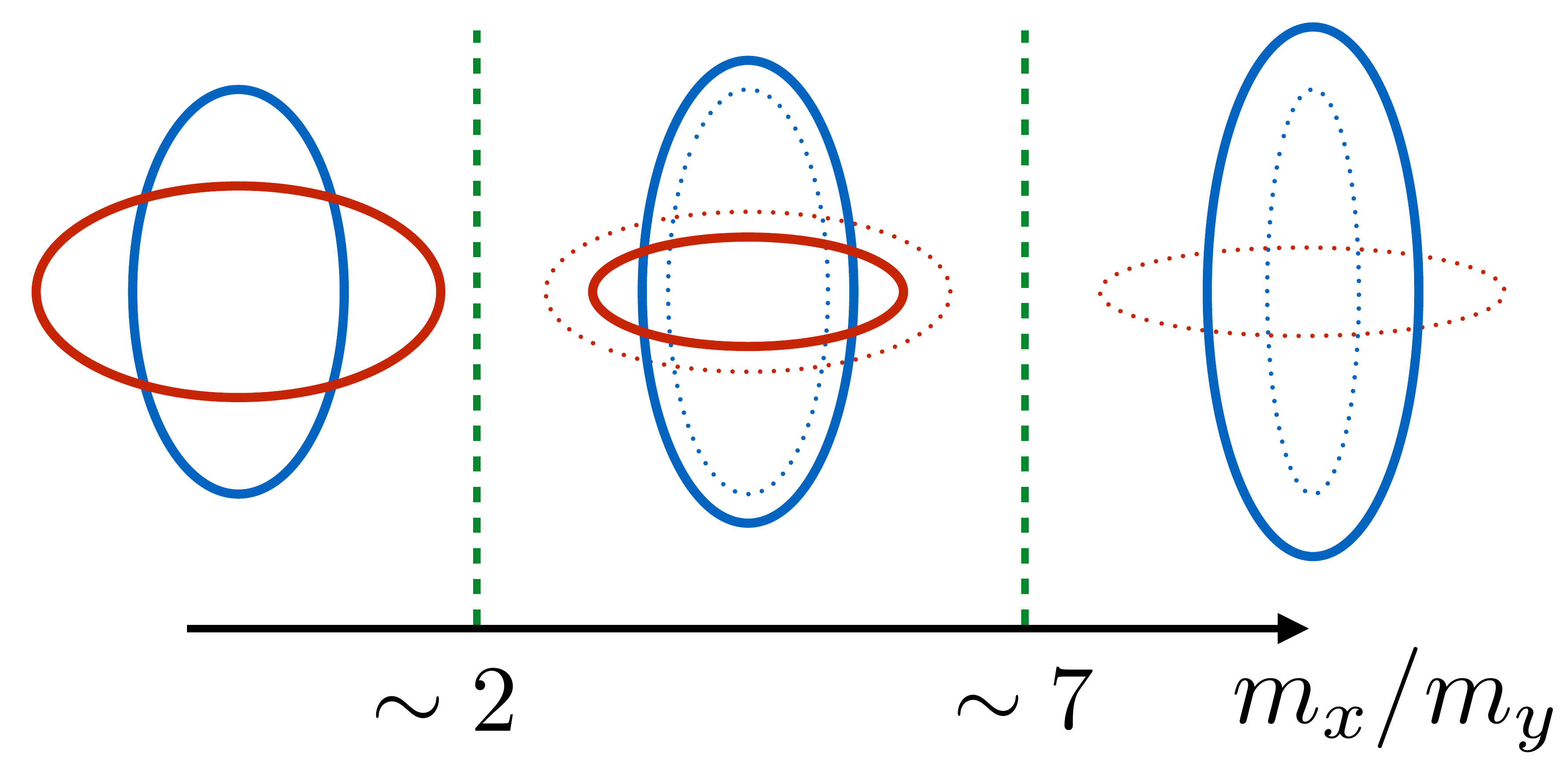}
\end{center}
\par
\renewcommand{\figurename}{Fig.}
\caption{(Color online) Schematic of the Stoner phase transition for a two-component CFL state. The solid blue and red ellipses are the composite Fermi surfaces in the two valleys. As a function of mass anisotropy the system first transitions into a partially polarized CFL state and finally into a fully polarized CFL state.}
\label{schematic}
\end{figure}

\begin{figure}[btp]
\begin{center}
\includegraphics[width=0.48\textwidth]{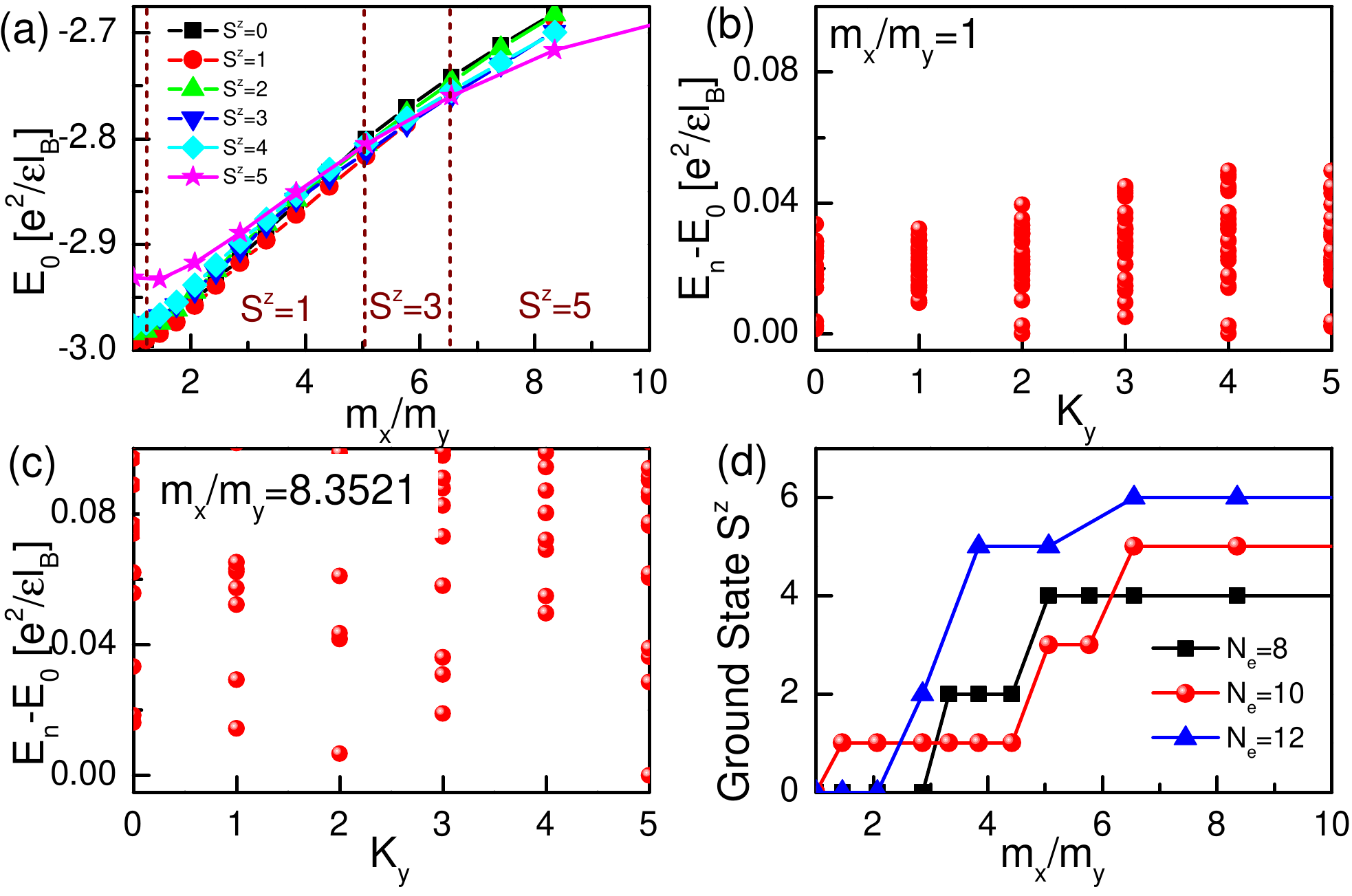}
\end{center}
\par
\renewcommand{\figurename}{Fig.}
\caption{(Color online) (a) The lowest energy in each pseudospin $S^z$ sector as a function of mass ratio for $n=0LL$ with $N_e=10$. The ground state at $m_x/m_y<1.3$ has $S^z=0$.  (b) and (c) are the energy spectra as a function of $K_y$ for unpolarized state and fully polarized state, respectively.  (d) DMRG results of  ground state $S^z$  as a function of mass ratio for $N_e=8,10,12$ systems. All of the electrons spontaneously polarize into a single flavor beyond a critical mass ratio $m_x/m_y \sim 7$.}
\label{Fig_Numerical}
\end{figure}

Mass anisotropy is expected to couple to concealed geometric degrees of freedom~\cite{Haldane2011} in FQH systems and can be experimentally tuned e.g. by  applying in-plane fields~\cite{ExpBx} or strain~\cite{Rokhinson2011,Jo2017}. Crucially, to this date numerical studies in the lowest Landau level ($n=0$LL) have demonstrated that the single component composite Fermi liquid (CFL) has a remarkable robustness against mass anisotropy~\cite{Ippoliti2017}. As we will demonstrate, however, a two-component half-filled $n=0$LL with anisotropic mass tensors rotated by $\pi/2$, as realized in AlAs quantum wells~\cite{Shayegan2006,Gokmen2010}, will undergo a quantum phase transition from an unpolarized two-component CFL into an analog of the Stoner ferromagnetic CFL state as a function of mass anistropy. We find specifically a transition from an unpolarized CFL into a partially polarized CFL at $m_x/m_y\sim2$, and finally into a fully valley polarized CFL at a critical value of the mass anisotropy $m_x/m_y \sim 7$ (see Figs.~\ref{schematic} and~\ref{Fig_Numerical}). AlAs has mass anisotropy of $m_x/m_y\sim5$ and therefore our results indicate that its CFL state is the analog of an itinerant Stoner magnet that is relatively close to the value for full valley polarization.

\section{Multicomponent anisotropic CFL.}   The Laughlin wavefunction~\cite{Laughlin83} is extremely good at minimizing the energy of a large class of repulsive interactions. The key to its energetic success is that it has $m$ zeros (or vortices) ``bound" to each particle. It describes a liquid at filling $\nu=1/m$ and is antisymmetric (symmetric) under coordinate exchange when $m$ is odd (even). The CFL trial wave-function~\cite{RR1994} can be viewed as a simple replacement to the Laughlin state at $\nu=1/2$ that corrects for its bosonic statistics while minimally altering its energetic virtues. As argued by Read~\cite{Read1994}, this is achieved by displacing the zeros of the bosonic Laughlin state by the smallest allowed amounts. As proposed in Ref.~\cite{RH2000}, this can be conveniently implemented in the torus leading to the following single component CLF trial wavefunction\cite{Apendix1}:
\be
|\Psi_{\text{CFL}}(\{{\bf d}_i\})\rangle=\det [\hat{t}_j({\bf d}_i)]| \Phi^{\text{Bose}}_{1/2}\rangle.
\ee
\noindent Here $| \Phi^{\text{Bose}}_{1/2}\rangle$ is the Laughlin wavefunction for $N_e$ bosons, $\hat{t}_j({\bf d})$ is the magnetic translation operator acting on particle $j$ by an amount ${\bf d}$, and $\{{\bf d}_i\}$ is a set of $N_e$ 2D vectors that parameterize the trial state. The many-body operator $\det [\hat{t}_j({\bf d}_i)]$ is antisymmetric under the exchange of the labels of any two particles and hence it maps bosonic states into fermionic ones. The vectors $\{{\bf d}_i\}$ are the variational parameters which control the displacement of zeros away from the particles, and, hence, energetically it is favorable for them to have the smallest possible magnitude. In order for the antisymmetrized operator not to vanish identically the ${\bf d}_i$ must all be distinct. On a torus only a discrete set magnetic translations are allowed~\cite{Haldane1985}, dictating that the displacement vectors $\{{\bf d}_i\}$ must be drawn to the following lattice:
\be\label{dipole}
{\bf d} \in \frac{m_1 {\bf L_1}+m_2 {\bf L_2}}{N_\phi}, \, m_{1,2}\in \mathbb{Z}{\rm mod}(N_\phi),
\ee
\noindent where ${\bf L_{1,2}}$ are the principal vectors of the torus and $N_\phi$ the number of flux quanta piercing it. The particles bound to displaced zeros (vortices) can be viewed as charge neutral dipolar objects with momenta ${\bf k}_i\equiv  l^{-2} \hat{{\bf z}}\times {\bf d}_i$~\cite{Read1994,RH2000,ChongSenthil,Hermanns,Jain2017}. These dipoles are the composite fermions and the disk-shaped region that they occupy in the lattice~\eqref{dipole} is their emergent fermi surface. This parametrization of the CFL has redundancies which can be viewed as ${\it gauge}$ degrees of freedom. In particular, changing the overall origin of the dipole lattice, and hence the origin of momentum, leads to the same physical state. A simple model for the energy of this state as a function of the set $\{{\bf d}_i\}$ or equivalently $\{{\bf k}_i\}$ can be written as~\cite{Shao2015}:
\begin{equation}\label{Evar}
\begin{split}
E[\{{\bf k}_i\}] \approx &\ E_0+\frac{1}{N_e} \sum_{i<j} \frac{|{\bf k}_i-{\bf k}_j|^2}{2 m^*}.\\
\end{split}
\end{equation}
\noindent Where, $m^*$, is the effective mass of the composite fermions. One of the great successes of this wavefunction is that it correctly captures the quantum numbers of the ground states and a few excited states even in systems with a relatively small number of particles~\cite{RH2000}. In particular, a quantum number that will play a crucial role in the analysis of our numerical results is the many body momentum, that is simply given by:
\be\label{manybK}
{\bf K}= \sum_{i}{\bf k}_i=2\pi \left(-\frac{\sum_{i}m_{2i}}{L_1} ,\frac{\sum_{i}m_{1i}}{L_2}\right),
\ee
\noindent here $(m_{1i},m_{2i})$ are the integers describing the dipolar lattice in Eq.\eqref{dipole} and the sums are defined modulo $N_e$~\cite{note1}.

We will now introduce a natural generalization of the CFL wavefunction to our case of a two valley system allowing for mass anisotropy. We label the two valleys by a pseudospin index $\sigma\in \{\uparrow, \downarrow\}$. The CFL wavefunction reads as follows:
\be\label{CFL3}
|\Psi_{\text{CFL}}(\{{\bf d_i^\uparrow}, {\bf d_i^\downarrow}\})\rangle=\det (\hat{t}_j({\bf d_i^\uparrow})) \det (\hat{t}_j({\bf d_i^\downarrow})) | \Phi^{\text{Bose}}_{1/2}\rangle.
\ee
\noindent Here $| \Phi^{\text{Bose}}_{1/2}\rangle$ is a two component Bosonic Laughlin wavefunction at total filling $\nu=1/2$ known as the Halperin $222$ state. We allow for the number of particles in each flavor to be a variational parameter. The state is here parametrized by two sets of vectors $\{{\bf d}^\sigma_i\}$ for $\sigma\in \{\uparrow, \downarrow\}$, drawn from the lattice in Eq.~\eqref{dipole}. The shapes of the lattice points that constitute the trial fermi surface of each component is allowed to be distinct to account for the possibility of fermi surfaces with different anisotropy for each valley. The relation between displacement and momenta is still given by ${\bf d}^\sigma_i=- l^{2} \hat{{\bf z}}\times {\bf k}_i^{\sigma}$, and the many-body momentum will given by the analogous expression to Eq.~\eqref{manybK} with a summation over both valleys $\sigma\in \{\uparrow, \downarrow\}$.

\begin{figure*}[tbp ]
\begin{center}
\includegraphics[width=0.8\textwidth]{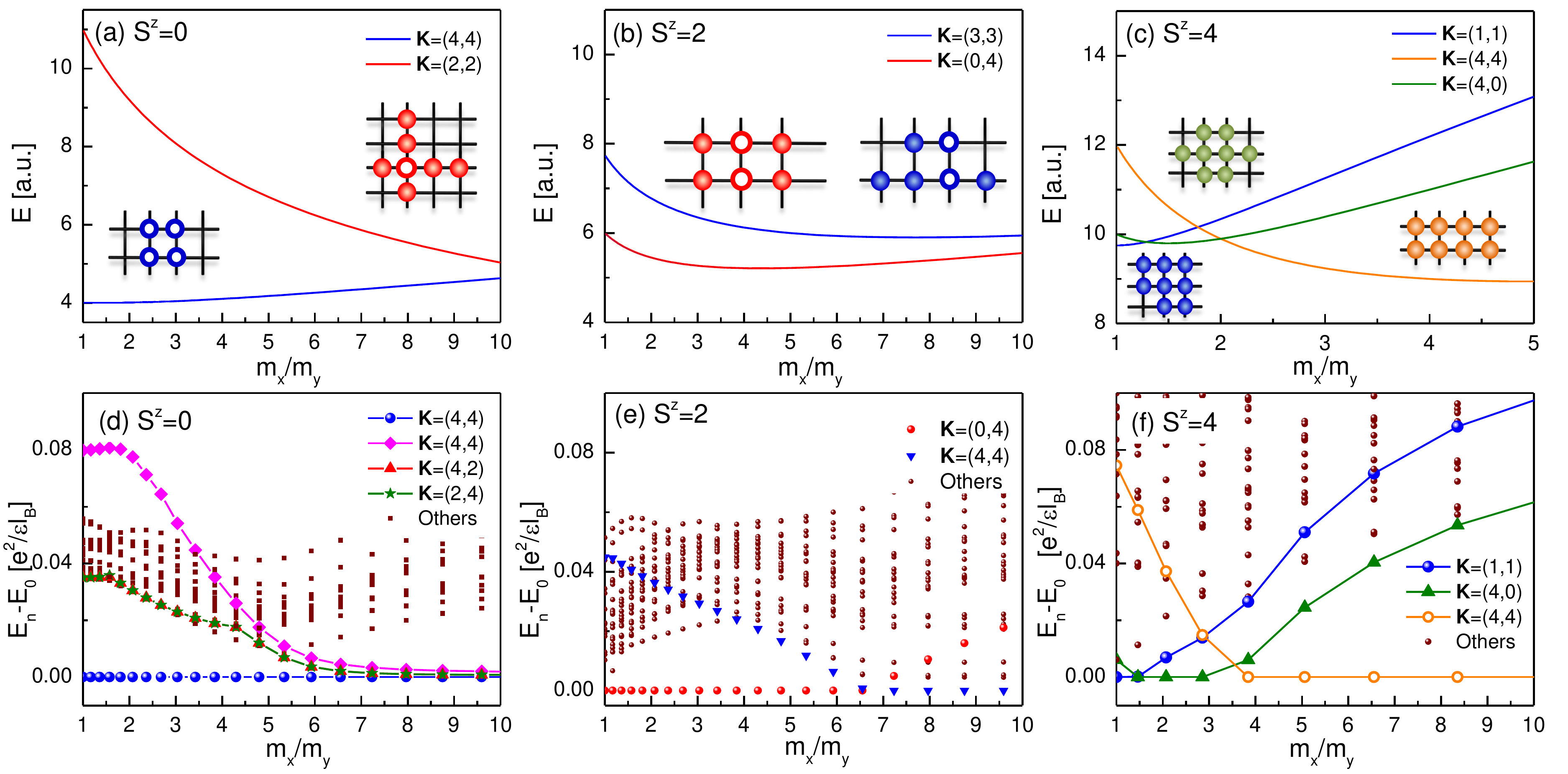}
\end{center}
\par
\renewcommand{\figurename}{Fig.}
\caption{(Color online) The comparison of the energy spectra of predicted many-body momentum sectors [(a) to (c)] with ED results [(d) to (f)] as a function of mass anisotropy. The theoretical predictions are consistent with numerics.}
\label{Ne=8Spectrum}
\end{figure*}

\section{Numerical solutions vs trial wavefunctions.}
Our model consists of electrons in two valleys with a kinetic energy of $ {H_0} = \frac{1}{{2m}}{g^{ab}_{m_\alpha}}{\Pi _a}{\Pi _b}$, with ${\Pi _a} = {p_a} - \frac{e}{c}{A_a}$ ($a,b=x,y$) and $\alpha=1,2$ are valley  indices. We choose $r\equiv\sqrt {m_y/m_x}$ to denote the mass anisotropy, then $g_{m_1}$ and $g_{m_2}$ are given by $g_{m_1} =\text{diag}[r,1/r]$ and $g_{m_2} =\text{diag}[1/r,r]$, respectively, corresponding to two perpendicular elliptical Fermi surfaces. We focus on total filling $\nu_T=1/2$ of the  $n=0$LL. In addition, the Hamiltonian includes the projected Coulomb interaction into the $n=0$LL, which reads as:
 \begin{align}\label {Ham}
V= \frac{1}{2 \pi N_\phi}\sum\limits_{i < j}  \sum\limits_{\bf{q},\bf{q} \ne 0} \sum\limits_{\alpha,\beta}  V (q) {e^{ - (q^2_{m_\alpha} +q^2_{m_\beta} )/4}}    {e^{i{\bf{q}} \cdot \left( {\bf{R}_{\alpha,i} - \bf{R}_{\beta,j}} \right)}} .
 \end{align}
\noindent Here,  $\alpha (\beta)$ is the valley index and  $V (q)= 2 \pi e^2 /q$ is the Fourier transformation of the Coulomb interaction. $q^2_{m_\alpha}=g^{ab}_{m_\alpha}q_aq_b$ includes the metric $g^{ab}_{m_\alpha}$ derived from the band mass tensor. Notice that the projected inter-valley interactions have full rotational invariance unlike the intra-valley interactions which have invariance only under $\pi$ rotations.

Before comparing our model with numerics, we analyze the symmetries. First, the system has separate conservation for the number of electrons in each valley due to the absence of inter-component tunneling, which allows to label eigenstates by pseudo-spin $S^z$. Second, there is also space symmetries like mirror times time reversal operation that act on the many body momenta. However, due to the mass anisotropy, only the subgroup of $\pi$ rotations (for a square torus), which map $(K_x,K_y)\rightarrow (-K_x,-K_y)$, remains a symmetry. Third, there is an interesting discrete symmetry which is the $\pi/2$ rotation combined with a pseudo-spin reversal operation $\uparrow  \leftrightarrow \downarrow$. This operation generally maps a state with $(K_x,K_y,S^z) \rightarrow (K_y,-K_x,-S^z)$, and therefore acts in a similar way to a space symmetry within the subspace $S^z=0$.

We combine exact diagonalization (ED) with density matrix renormalization group (DMRG) methods on a square torus~\cite{Haldane1985,Yoshioka1984} . Figure~\ref{Fig_Numerical} (a) shows the lowest energy in each pseudo-spin $S^z\equiv (N_e^\uparrow-N_e^\downarrow)/2$ sector, as a function of the mass anisotropy for $N_e=10$ system. We find a transition from a valley un-polarized state at small mass anisotropy into a partially polarized state and finally transitions into a fully polarized state at a higher critical mass anisotropy~\cite{note2}.
{Fig.~\ref{Fig_Numerical} (b) and (c) shows the spectrum as a function of momenta for a representative values of the mass anisotropy, and we see that there is no clear gap in the spectrum indicating that both states are gapless. To be able to reach larger system sizes we resort to DMRG~\cite{FeiguinDMRG}. Figure~\ref{Fig_Numerical} (d) displays the ground state  $S^z$  as a function of mass anisotropy obtained from DMRG, which is in agreement with the results at smaller system sizes obtained in ED. From these results we estimate that the system is able to transition into a fully polarized state at a  critical mass anisotropy of $m_x/m_y \sim 7$.  We hope future studies will access larger sizes to evaluate the persistence of the trends we have found and estimate more precisely the phase transition boundaries and their nature closer the thermodynamic limit.

To compare with CFL trial wavefunctions, we introduce the following generalization of Eq.~\eqref{Evar} to capture the energetics of these states:
\begin{equation}\label{updown}
\begin{split}
& E[\{{\bf k}_i^{\sigma}\}] =\sum_{\sigma} \sum_{i<j}^{N_e^\sigma} \epsilon_\sigma ({\bf k}_i^{\sigma}-{\bf k}_j^{\sigma}) +\sum_{i \in \uparrow,j \in \downarrow} \epsilon_{\uparrow\downarrow} ({\bf k}_i^{\uparrow}-{\bf k}_j^{\downarrow}),\\
& \epsilon_\uparrow ({\bf k}) \approx \frac{k_x^2}{2 m_x^* N_e}+\frac{k_y^2}{2 m_y^* N_e}, \ \epsilon_{\uparrow\downarrow} ({\bf k}) \approx \frac{|{\bf k}|^2}{2 m_{\uparrow\downarrow}^* N_e}. \\
\end{split}
\end{equation}
\noindent and $\epsilon_\downarrow$ is obtained from $\epsilon_\uparrow$ by changing $m_x^* \leftrightarrow m_y^*$. Upon projecting into a Landau level the composite fermions kinetic energy arises entirely from the microscopic interactions and hence their symmetries follow those of the interaction terms. This motivates the isotropy of the inter-flavor kinetic term, which follows from the isotropy of the interflavor interaction captured by the form factors in Eq.~\eqref{Ham}~\cite{note3}.
The key predictions of this model that we will contrast against numerical simulations are the many body momentum sectors ${\bf K}$ of the ground state expected as a function of mass anisotropies.

Let us consider the system with $N_e=8$ particles.  We start with studying the fully polarized sector (i.e., $S_z=4$), which corresponds to the single component CFL. The momentum sector of ground state predicted from Eqs.~\eqref{updown} as a function of mass anisotropy, is depicted in Fig.~\ref{Ne=8Spectrum} (c). The momentum sectors predicted are in perfect agreement with the results from ED shown in Fig.~\ref{Ne=8Spectrum} (f). We consider next partially polarized states. Motivated by the scaling of the form factors in the lowest Landau level and by previous numerical findings on the scaling of the effective composite fermion mass as a function of the bare mass~\cite{Ippoliti2017}, we have chosen the following phenomenological parametrization of the effective masses: $m_{\uparrow x} = m^* \sqrt{m_x/m_y}, m_{\uparrow y} = m^* \sqrt{m_y/m_x}, m^*_{\uparrow\downarrow}=  (m_{\uparrow x}+m_{\uparrow y})/2$.
\noindent Using this simple model we obtain that the ground states at $S_z=0$ and $S_z=2$ are expected to have many body momenta in the sectors $(4,4)$ and $(0,4)$ in units of $2\pi/L$, respectively [see Fig.~\ref{Ne=8Spectrum} (a) and (b)]. On the other hand the numerical results are shown in Fig.~\ref{Ne=8Spectrum} (d) and (e). In the case of $S_z=2$ we see a level crossing which appears in  ED is not captured in the simple model, however such level crossing occurs when the $S_z=2$ sector is not the absolute ground state sector [see Fig~\ref{Fig_Numerical} (d)]. Therefore,  we conclude that the quantum numbers of the ground state from this simple model always coincide with those seen in ED whenever the corresponding $S_z$ is the absolute ground state sector ~\cite{note4}.

\begin{figure}[btp]
\begin{center}
\includegraphics[width=0.4\textwidth]{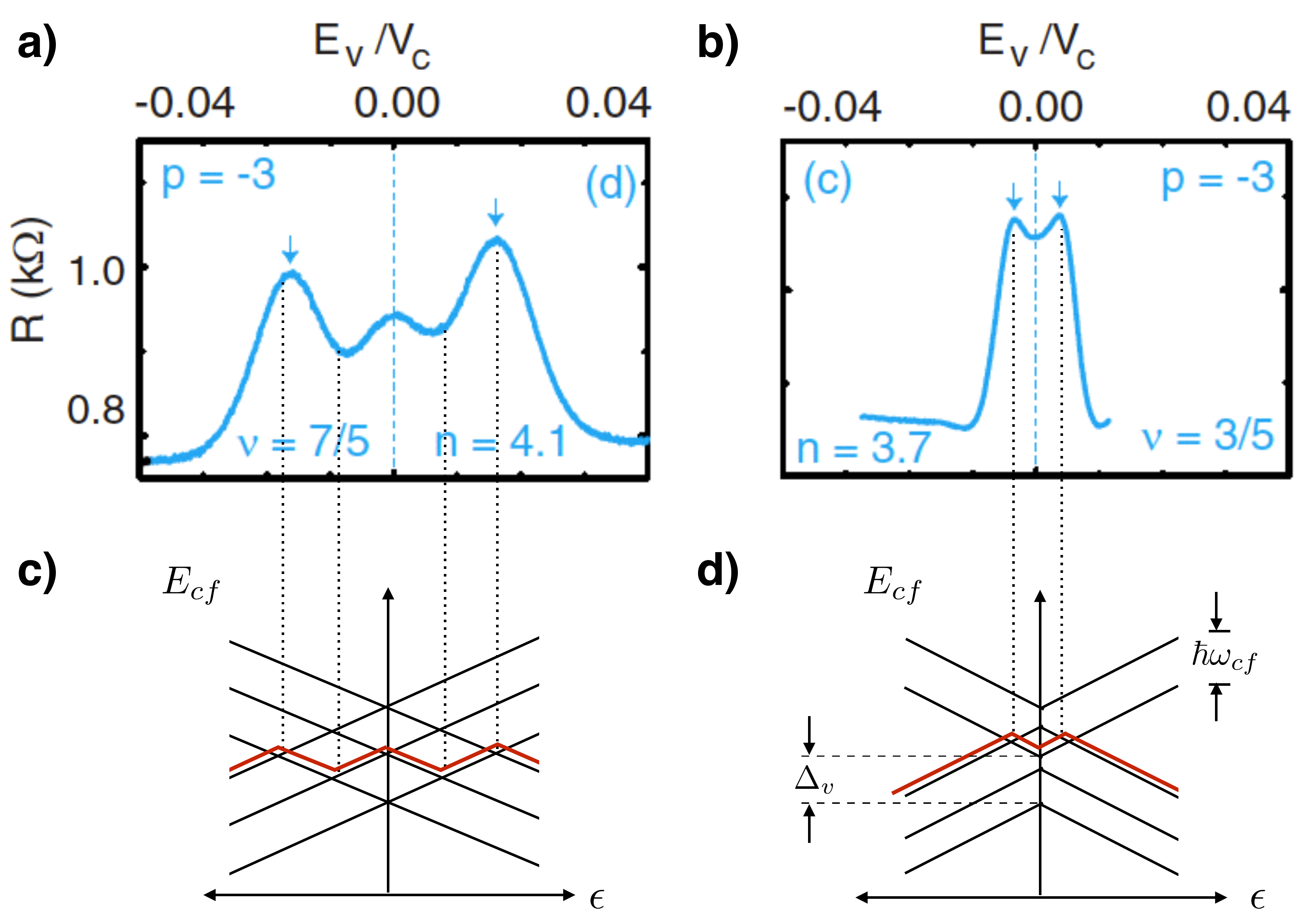}
\end{center}
\par
\renewcommand{\figurename}{Fig.}
\caption{Resistance traces of the $p=-3$ Jain states surrounding $\nu=3/2$ and $\nu=1/2$ as a function of strain-induced valley splitting from Ref.~\onlinecite{Shayegan-ph}. The trace near $\nu=3/2$ in panel (a) can be understood assuming that the spontaneous valley splitting $\Delta_v$, if present, is smaller than the effective cyclotron energy of composite fermions $\hbar \omega_c$. However, we propose that the resistance trace surrounding $\nu=1/2$ in panel (b) implies an spontaneous valley splitting that is larger than the composite fermion effective cyclotron splitting.}
\label{cfcrossing}
\end{figure}

\section{Connections to experiments.} In AlAs quantum wells~\cite{Gokmen2010,gfactor,Vakili,Gokmen2,LLmixing,Shayegan-ph},  the mass anisotropy is $m_x/m_y \sim 5$. Our findings indicate that its CFL at $\nu_T=1/2$ will likely be a Stoner ferromagnetic state with partial valley polarization~[see Fig.~\ref{Fig_Numerical}(d)]. As we will argue next, there are clear hints of this behavior in existing experimental studies of AlAs.

One of the most clear hints is offered by comparing the FQH states in the Jain sequence that surrounds $\nu=1/2$ with those at $\nu=3/2$~\cite{Shayegan-ph,Bishop}. In a two-component system in which there is no spontaneous polarization of the CFL, such as our system in the limit $m_x/m_y=1$, it is well known that the Jain sequence is expected to also follow this trend and the composite fermions tend to form FQH states with minimal valley polarization~\cite{JainBook}. In particular this implies that the Jain states with even numerators will be valley unpolarized. As a function of the single particle valley splitting, which is controlled by straining the sample, one therefore expects an analogous pattern of composite fermion level crossings as those of the non-interacting Landau levels of free fermions as a function of strain. This trend is indeed seen in the piezoresistance traces for the fractions that surround the composite fermi sea at $\nu=3/2$~\cite{Bishop,Shayegan-ph}. However, for the fractions surrounding $\nu=1/2$ deviations have been detected. In particular the trace for $\nu=3/5$ displays a  minimum near zero strain separated by a single maximum from the saturated state (see Fig.~\ref{cfcrossing}(b)), in contrast to its particle hole conjugate at $\nu=7/5$, which displays a maximum near zero strain separated by another maximum at finite strain before the saturation of the resistivity is reached (see Fig.~\ref{cfcrossing}(a)). This pattern can be explained if one assumes that there is an spontaneous valley splitting, $\Delta_v$, at $\nu=3/5$, which is larger than the composite fermion effective cyclotron energy so that only one level crossing is realized as a function of strain as depicted in Fig.~\ref{cfcrossing}(d). This offers a clear indication that the system is proximate to the Stoner instability and presumably this tendency is enhanced as the Landau level mixing is reduced, which explains the enhanced tendency at $\nu=1/2$ compared to $\nu=3/2$.

Another hint of Stoner magnetism near $\nu=1/2$ is the reduced value for the critical strain needed to achieve saturation of the resistivity as compared to $\nu=3/2$~\cite{Shayegan-ph,Shayegan-piezo}, which can be interpreted as an enhanced susceptibility to induce full valley polarization. The number of FQH states observed in AlAs is rather limited, but we hope that our study will motivate the realization of samples with fractions closer to half-filling where the trends might offer a closer look into the physics of the composite fermi sea state. We also hope that our study motives future theoretical explorations of the role of aspects beyond the ideal model we have considered, in particular, on the impact of Landau level mixing.

\section{Discussion and Summary}.
The Stoner transition that we have encountered is accompanied by spontaneous breaking of discrete space symmetries. Specifically, the $\pi/2$ rotation composed with valley exchange is spontaneously broken. The remaining rotational symmetry is a $\pi$ rotation. Thus, from the symmetry point of view, our state has an Ising nematic character analogous to the quantum Hall ferromagnet at $\nu=1$~\cite{Abanin,ferro}. However, it is crucially different in that it is an itinerant system with a gapless Fermi surface unlike the state at $\nu=1$ which has a fully gapped bulk spectrum. Our Ising nematic Stoner transition also differs from a pomeranchuk Ising nematic transition (such as those recently studied numerically in Refs.~\cite{Sam,Ashvin,Lee2018}), because in our case the broken symmetry is not a pure space symmetry but includes a composition with a pseudospin operation~\cite{note5}.
It is also differs from other higher angular momentum channel instabilities such as those considered in Refs. \cite{Fradkin}.

We would like to comment on other proposals for the realization of Stoner-like CFLs which are distinct from ours. Ref.~\cite{JainStoner} proposed a Stoner instability for isotropic CFLs of higher order flux at lower filling fractions supported by trial wave-function calculations. Another proposal is the interlayer coherent CFL phase in quantum Hall bilayers~\cite{Alicea}. To the best of our knowledge, there is no experimental support for the realization of neither of these proposals. More recently Stoner magnetism in the $n=1$LL of graphene was put forth in Ref\cite{Balram2015}. Experiments \cite{Amet2015} have uncovered several incompressible states in the $n=1$LL of graphene, but, there is no direct experimental evidence as of yet for the proposed Stoner state of composite fermions to best of our knowledge.

Our results are consistent with the fact that as a function of mass anisotropy the two-component half-filled $n=0$ LL of a system with the symmetry of AlAs undergoes an spontaneous polarization transition from a composite fermi sea state of equal occupation of both flavors into one in which a single flavor is occupied as the mass anisotropy is increased, in a manner analogous to the Stoner picture of magnetism of itinerant magnets. This conclusion is drawn on the remarkably good agreement of a simple model based on two composite fermion fermi seas to reproduce the quantum numbers of finite size systems in the torus, for both the fully polarized state in which the composite fermions have an elliptically shaped fermi surface as well as partially polarized cases. We note that for all the system sizes that we have reached we find that the un-polarized state transitions through a series of intermediate states with partial polarization and we have found evidence that the intermediate states with partial polarization have quantum numbers that can be explained within the picture of two composite fermi surfaces of different sizes for each flavor. Additionally, we have found experimental evidence of the physics of Stoner magnetism in the Jain states surrounding $\nu=1/2$ in AlAs quantum wells, lending support to the belief that our findings remain robust after including more realistic corrections to our ideal model. We also hope that our study motives future theoretical explorations of the role of aspects beyond the ideal model we have considered, in particular, on the impact of Landau level mixing.

\begin{acknowledgments}
We would like to thank Mansour Shayegan, Jainendra Jain and Eduardo Fradkin  for valuable discussions. Z.Z.  and L.F. are supported by the David and Lucile Packard foundation. D.N. Sheng is supported by the U.S. Department of Energy, Office of Basic Energy Sciences under grants No. DE-FG02-06ER46305.
\end{acknowledgments}

\appendix
\renewcommand{\theequation}{S\arabic{equation}}
\setcounter{equation}{0}
\renewcommand{\thefigure}{S\arabic{figure}}
\setcounter{figure}{0}

\section {Anisotropic single-component composite fermi seas}
In this section, we briefly review the single component CFL trial states. The classic wavefunction describing the composite fermi sea in the half-filled Landau level was introduced by Read and Rezayi~\cite{RR1994}, motivated by the field theoretic description of Halperin, Lee, and Read~\cite{HLR} and as a natural generalization of the fractional quantum Hall Jain wave-functions~\cite{JainBook}. The wavefunction can be written as:
\be
\Psi_{\text{CFL}}=P_{\text{LLL}}[\Psi_{\text{FL}}\Phi^{\text{Bose}}_{1/2}],
\ee
\noindent where $\Phi^{\text{Bose}}_{1/2}$ is the bosonic Laughlin wavefunction at total filling $\nu=1/2$ and $\Psi_{\text{FL}}$ is the Slater determinant of a non-interacting Fermi gas at zero magnetic field. $P_{\text{LLL}}$ projects the wavefunction onto the lowest Landau level. Explicit evaluations involving this wavefunction are commonly done in the disk or sphere geometry as it is non-trivial to perform the projection in the torus geometry (see however~\cite{Hermanns, Jain2017}). A convenient way to write the composite fermi liquid state on the torus was introduced by Rezayi and Haldane~\cite{RH2000}:
\be
|\Psi_{\text{CFL}}(\{{\bf k}_i\})\rangle=\det (e^{i {\bf k}_i \cdot \hat{{\bf R}}_j})| \Phi^{\text{Bose}}_{1/2}\rangle.
\ee
\noindent Here, the Bosonic Laughlin wavefunction is viewed as a state $| \Phi^{\text{Bose}}_{1/2}\rangle$ of $N_e$ bosons, upon which the operator $\det (e^{i {\bf k}_i \cdot \hat{{\bf R}}_j})$ acts. $\{{\bf k}_i\}$ is a set of $N_e$ distinct 2D vectors that parameterize the state and $\hat{{\bf R}}_j$ is the guiding center operator of particle $j$. Because the guiding center operators are intra-Landau level operators this wavefunction is manifestly in the Lowest Landau level, and the antisymmetrization of the determinant guarantees that the state $|\Psi_{\text{CFL}}(\{{\bf k}_i\})\rangle$ describes fermions.

To gain physical insight into this wavefunction, it is convenient to replace guiding center operators in favor of magnetic momentum operators, which are the generators of the intra-Landau level magnetic translation algebra. The magnetic momentum, $\hat{{\bf Q}}$, is a rotated and rescaled version of the guiding center operator: $\hat{{\bf R}}=l^{2} \hat{{\bf z}}\times \hat{{\bf Q}}$. Using this relation the wavefunction can be rewritten as the action of non-trivial translation operator on the particles of the bosonic Laughlin state
\be\label{CFL2}
|\Psi_{\text{CFL}}(\{{\bf d}_i\})\rangle=\det (\hat{t}_j({\bf d}_i)) | \Phi^{\text{Bose}}_{1/2}\rangle, \ {\bf d}_i\equiv - l^{2} \hat{{\bf z}}\times {\bf k}_i,
\ee
\noindent where $\hat{t}_j({\bf d}_i)=e^{i {\bf d}_i \cdot \hat{{\bf Q}}_j}$ is a magnetic translation operator acting on particle $j$ by an amount ${\bf d}_i$. Equation~\eqref{CFL2} provides an appealing picture of the composite fermi sea as a quantum liquid of fermionic dipoles, first advocated by Read~\cite{Read1994}. When the bosonic state, $\Psi_{1/2}$, is viewed as a function of a single coordinate ${\bf z}_i$ while other coordinates are held fixed, ${\bf z}_i$ has a zero of degree 2 at any other particle position ${\bf z}_j$. After the action of $\det (t_j({\bf d}_i))$, the resulting function of ${\bf z}_i$ has no longer the zeroes at ${\bf z}_j$ but instead at ${\bf z}_j+{\bf d}_j$. Thus this wavefunction is in a sense one in which the zeroes are displaced relative to the bosonic state in such a way so that the state is fully antisymmetric~\cite{note}.
It is clear that the trial state with the minimal energy would be that in which the set of $\{{\bf d}_i\}$ have minimal magnitude and hence are distributed in a disk around the origin. This is the energetic origin of the fermi sea state. On this basis, a simple model for the energy of this state as a function of the set $\{{\bf d}_i\}$ or equivalently $\{{\bf k}_i\}$ can be written as~\cite{Shao2015}:
\begin{equation}\label{xx}
\begin{split}
E[\{{\bf k}_i\}] =& \ E_0+\frac{1}{N_e} \sum_{i<j} \epsilon ({\bf k}_i-{\bf k}_j) \\
\approx & \ E_0+\frac{1}{N_e} \sum_{i<j} \frac{|{\bf k}_i-{\bf k}_j|^2}{2 m^*},\\
\end{split}
\end{equation}
\noindent here $E_0$ is an energy on the order of the Bosonic Laughlin state ground state energy, which would be the energy of a state in which the statistics allow all the particles to condense into a common ${\bf k}$. When the system is placed on a torus only a discrete subset of single particle magnetic translations are compatible with fixed boundary conditions~\cite{Haldane1985}. This implies that the displacement vectors belong to a lattice
\be\label{dipole}
{\bf d} \in \frac{m_1 {\bf L_1}+m_2 {\bf L_2}}{N_\phi}, \, m_{1,2}\in \mathbb{Z}{\rm mod}(N_\phi),
\ee
\noindent where ${\bf L_{1,2}}$ are the principal vectors of the torus. Because of its high symmetry the Bosonic Laughlin state forms at zero many body momentum, and therefore the many-body momentum of the corresponding composite fermi sea state is determined completely by the displacement wave-vectors. For simplicity we consider a rectangular torus. Then, the many body momentum of the CFL state will be
\be\label{manybK}
{\bf K}=  \frac{\hat{{\bf z}}}{l^2}\times \sum_{i}{\bf d}_i=2\pi \left(-\frac{\sum_{i}m_{2i}}{L_1} ,\frac{\sum_{i}m_{1i}}{L_2}\right),
\ee
\noindent where $(m_{1i},m_{2i})$ are the integers describing the dipolar displacement of particle $i$ from Eq.~\eqref{dipole}, and the sums are defined modulo $N_e$~\cite{note1}
Notice that this description of the fermi liquid state has redundancies which can be viewed as ${\it gauge}$ degrees of freedom. In particular, changing the overall origin the dipole lattice, and hence the origin of momentum, leads to the same physical state.


\begin{thebibliography}{99}
\bibitem{note1}This is for the case of half-filling $\nu=1/2$. More generally at filling $\nu=p/q$ where $p,q$ have no common divisor greater than $1$, the sums are defined modulo $N$, where $N_\phi=q N$~\cite{Haldane1985}.
\bibitem{note2}The spin of the $N_e=10$ system has $S_z=1$ even at small mass anisotropies. We believe that this is a finite size effect leading to a fluctuation of order $1$ in $S_z$ which arises from the absence of a pseudo-spin flip gap combined with the inability to form a simple closed shell in momentum space for small mass anisotropies in the $Sz=0$ sector for this specific system size.
\bibitem{note3}In higher Landau levels this isotropy will reduce to a $C_4$ symmetry, so this form is expected to capture the symmetries only to leading order in $k$.
\bibitem{note4}We have chosen to illustrate this with the results with 8 particles because the analysis is considerably simpler as this finite size state has a particularly simple closed shells, but we have verified that similar agreement is achieved for 10 particles as well.
\bibitem{note5}It is amusing to note that this distinction is only sharp for the ideal model we are considering in which there is no inter-valley swap or scattering terms but only density-density interactions. From a crystal symmetry point of view, the valleys in AlAs are exchanged by a $C_4$ rotation and therefore the real space symmetry breaking $C_4 \rightarrow C_2$, resembling the situation in models of metallic criticality. However, the model we are considering is believed to be an excellent description of AlAs deep in the quantum Hall regime for which the terms that would reduce the symmetry of the Hamiltonian can be safely neglected.
\bibitem{note} Notice that antisymmetry of the wave-function implies that at least one zero must coincide with the particle position, which motivates an alternative wavefunction obtained by multiplying the Fermi sea of bosons at $\nu=1$ by a Jastrow factor~\cite{ChongSenthil}, whether these two wafunctions describe the same phase of matter in the thermodynamic limit is a subject of current debate, but in practice for small finite size systems both wavefunctions are expected to display similar physics.
\end{thebibliography}
\end{document}